\documentclass[12pt,a4paper]{article}
\usepackage{amssymb}
\usepackage{a4,graphicx}
\begin{document}

\title{ Magnetic field effect on tunnel ionization of deep
impurities by terahertz radiation}
\author{S.D.~Ganichev$^{1,2}$, S.N.~Danilov$^{1}$,
M.~Sollinger$^{1}$,\\
J.~Zimmermann$^{1}$, A.S.~Moskalenko$^{2}$, V.I.~Perel$^{2}$,
I.N.~Yassievich$^{2}$,\\ C.~Back$^{1}$ and W.~Prettl$^{1}$\\
{\it  $^{1}$ Universit\"at Regensburg, D~93040 Regensburg, Germany}\\
{\it $^{2}$A.F.Ioffe Physico-Technical Institute, St. Petersburg,
Russia}
 }

\maketitle

\begin{abstract}
A suppression of tunnelling ionization of deep impurities in
terahertz frequency electric fields by a magnetic field is
observed. It is shown that the ionization probability at external
magnetic field, \textbf{\textit{B}}, oriented perpendicular to the
electric field of terahertz radiation, \textbf{\textit{E}}, is
substantially smaller than that at
$\textbf{\textit{B}}\parallel\textbf{\textit{E}}$. The effect
occurs at low temperatures and high magnetic fields.
\end{abstract}

\vspace{4cm}

\small
{Name and contact information of corresponding author:\\
S.D. Ganichev\\
Institut f\"ur Exp. und Angew. Physik\\
Universit\"at Regensburg\\
D--93040 Regensburg,
Germany\\
Phone: +49-(941) 943-2050, \hspace{0.3cm}
Fax: +49-(941) 943-4223\\
e-mail: sergey.ganichev@physik.uni-regensburg.de}

\clearpage

\section{Overview}

Tunneling ionization of deep centers in a terahertz field of
high-intensity far-infrared laser radiation, with photon energies
tens of times  lower than the impurity binding energies has been
investigated in great detail during the last
decade~\cite{review2003tun}. In contrast to tunneling ionization
of atoms, where only electron tunneling takes place, ionization of
impurities in solids is accomplished by two simultaneous
processes: electron tunneling through the potential well  formed
by the attractive force of the impurity  and the externally
applied electric field and  the redistribution of the vibrational
system by defect tunneling. At very high  radiation electric field
strengths direct tunneling may occur without involving phonons.

The tunneling probability is independent on the radiation
frequency up to very high frequencies. In this  quasi-static
regime, electrons tunnel at constant energy in a time much shorter
than the reciprocal radiation frequency $\omega^{-1}$. At higher
frequencies, however, tunneling probability drastically increases
in comparison  to the tunneling in $dc$ field of the same field
strength. In such a  high-frequency regime electrons can absorb
energy from the radiation field during tunneling leaving the
barrier at higher energy. By this the effective width of the
tunneling barrier is reduced and, thus, the tunneling probability
enhanced.

In a semiclassical approach defect tunneling takes place from the
adiabatic potential corresponding to  the impurity bound state to
the state where the carrier is detached from the impurity. The
tunneling probability is controlled by the B{\"u}ttiker-Landauer
tunneling time~\cite{Landauer1994p217}  $\tau$ which is a function
of temperature. The transition from frequency independent
tunneling in a classical electric field to fully quantized
multi-photon transitions occurs in the terahertz range and may be
explored applying high-power far-infrared lasers. The borderline
is given by $\omega\tau = 1$ where $\omega$ is the radiation
frequency.

These considerations are based on semiclassical theory where the
carriers have a classical trajectory. In this case the tunneling
probability is expected to be affected by the strength and the
orientation of an external magnetic field. For electron tunneling
through static potential barriers this effect was theoretically
investigated in~\cite{Kotova68p616} and observed in quantum well
structures~\cite{Eaves}. The theory has been extended for
phonon-assisted tunneling ionization of deep impurities in $dc$
electric  fields~\cite{Perel98p804} and in high frequency
alternating fields~\cite{physicaB99,Moskalenko00p217} showing that
also in the case of phonon-assisted tunneling, even in the
high-frequency regime, the carrier emission is suppressed by an
external magnetic field ($\textbf{\textit{B}}\perp
\textbf{\textit{E}}$). In this work we give evidence for the
applicability of the semiclassical model to tunneling assisted by
phonons which is concluded from the experimental observation of
the suppression of tunneling probability by an external magnetic
field oriented perpendicular to the carrier trajectory.
\section{Tunneling in alternating fields}
We have shown in~\cite{review2003tun} that the probability of
phonon assisted tunneling depends on the strength of
\textbf{\textit{E}}, an oscillating electric field of frequency
$\omega$:
\begin{equation}
   e (E) \propto \exp \left[ \frac{E^2}{(E_c^*)^2}\right] \quad
   {\rm with} \quad
   (E_c^*)^2 = \frac{3 m^* \hbar}{e^2 (\tau^*)^3}
   \label{E18}
\end{equation}
with electron effective mass $m^*$ and
\begin{equation}
   (\tau^*)^3 = \frac{3}{4 \omega^3} (\sinh (2 \, \omega \tau)
   - 2 \, \omega \tau)
   \label{E20}
\end{equation}
The frequency dependence of the tunneling process is controlled by
$\omega \tau$ where the tunneling time $\tau$ depends on
temperature: $\tau\,= \hbar/2kT\,\pm\tau_{1}$. Here
 $\tau_1$ is  of the order of the period of the impurity vibration and plus and minus
 correspond to substitutional and auto-localized impurities, respectively~\cite{PRL95}.
It follows from Eq.~(\ref{E20}) that for $\omega\tau \ll 1$ the
effective time $\tau^*$ gets equal to the tunneling time $\tau$
yielding a well known formula  of phonon assisted tunneling in the
quasi-static regime~\cite{review2003tun}.

In the presence of an external magnetic field oriented
perpendicular to the electric field of radiation  the functional
dependence of the probability on the electric field strength
remains unchanged, however the value of the effective time
$\tau^*$ becomes dependent on the magnetic field strength. An
increase of the cyclotron frequency $\omega_c = eB/m^* $ over the
reciprocal tunneling time results in the decrease of the tunneling
probability. The suppression of the tunneling probability occurs
in both frequency ranges, at low frequencies, when tunneling is
independent of radiation frequency, as well as at high frequencies
when the tunneling probability increases drastically with rising
frequency. The effect of a magnetic field \textbf{\textit{B}} on
tunneling is strongest if it is oriented normal to  the radiation
electric field  \textbf{\textit{E}} and vanishes if
$\textbf{\textit{B}}\parallel\textbf{\textit{E}}$. For the
ionization probability we again obtain an exponential dependence
on the square of the electric field strength in the form of
Eq.~(\ref{E18}), however, now the effective time $\tau^*$ depends
 on the magnetic field strength:

\begin{eqnarray}
(\tau^*)^3=\frac{3\omega_c^2}{(\omega^2-\omega_c^2)^2}
\left\{\int_0^{\tau}\left(-\cosh\omega\tau^\prime +
\frac{\omega_c}{\omega}\frac{\sinh\omega\tau}{\sinh\omega_c\tau}
\cosh\omega_c\tau^\prime\right)^2d\tau^\prime \right.\nonumber\\
\left.
+\int_0^{\tau}\left(\frac{\omega}{\omega_c}\sinh\omega\tau^\prime-
\frac{\omega_c}{\omega}\frac{\sinh\omega\tau}{\sinh\omega_c\tau}
\sinh\omega_c\tau^\prime\right)^2  d\tau^\prime \right\}.
\label{E50}
\end{eqnarray}
In Fig.~1 the calculated impurity ionization probability is
plotted as a function of $E^2$ for various magnetic field
strengths showing the drop of the carrier emission rate with
increasing \textbf{\textit{B}}. The data are presented for two
different values of $\omega \tau$ representing the quasi-static
and the high-frequency regimes.
%
\begin{figure}[!h]
\centering
\includegraphics[scale=0.3]{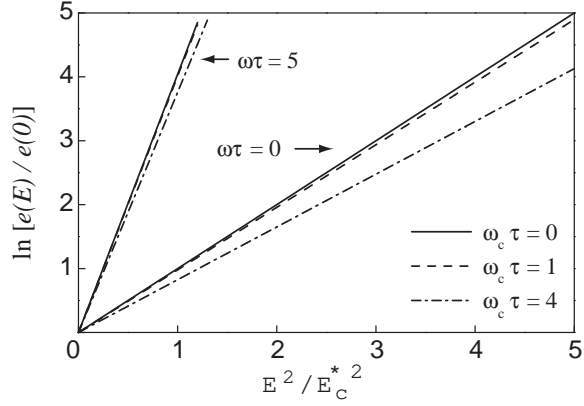}
 \caption{ln$[e(E)/e(0)]$ as a function of  $E^2/E_c^{*2}$ for
different $\omega\tau$ and $\omega_c\tau$ calculated after
Eqs.~\protect (\ref{E18}), (\protect \ref{E20}) and \protect
(\ref{E50}) for a magnetic field normal to the electric field;
$\omega$ and $\omega_c$ are the radiation frequency and the
cyclotron frequency, respectively. Here the parameter $\omega\tau$
controls the frequency dependence of tunneling while
$\omega_c\tau$ reflects the influence of the magnetic field.}
\label{F6}
\end{figure}
\section{Experimental Technique and Results}
The experiments were carried out on mercury doped germanium in the
temperature range of 10\,K to 70\,K. Tunneling ionization has been
achieved by far--infrared laser radiation with photon energies
much smaller than the thermal impurity ionization energy
$\varepsilon_T = 90$\,meV. The radiation source was a pulsed
far-infrared molecular laser optically pumped by a TEA--CO$_2$
laser. Operating the optically pumped laser with NH$_3$ and
CH$_3$F as active gases, 40~ns pulses with intensity up to 2
MW/cm$^2$ have been obtained at wavelengths of 148~$\mu$m and
496~$\mu$m. The ionization probability has been measured by
detecting photoconductivity~\cite{review2003tun}. To apply a small
probe voltage to the sample two ohmic contacts along $x$ direction
were prepared. An external magnetic field \textit{\textbf{B}} up
to 7.5~T has been applied along $x$. The ionization probability as
a function of the radiation intensity $I \propto E^2$ has been
obtain for $\textbf{\textit{E}}
\parallel B_x$ and $\textbf{\textit{E}} \perp
B_x$ for different magnetic field strengths. In addition the
magnetic field dependence of the signal has been determined for
constant intensity. The orientation between the external magnetic
field $B_x$ and the electric field \textbf{\textit{E}} of linearly
polarized terahertz radiation  has been  changed by means of a
$\lambda/2$ crystal quartz plate.

\begin{figure}[!t]
\centering
\includegraphics[scale=0.3]{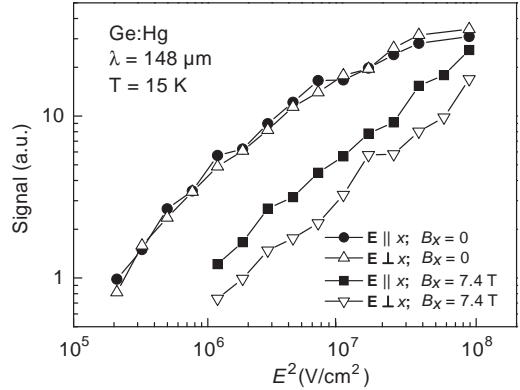}
\caption{Photoconductive signal  for Ge:Hg as a function of $E^2$
for different magnetic field strengths $B_x$ and orientations.  }
\label{F2}
\end{figure}

\begin{figure}
\centering
\includegraphics[scale=0.3]{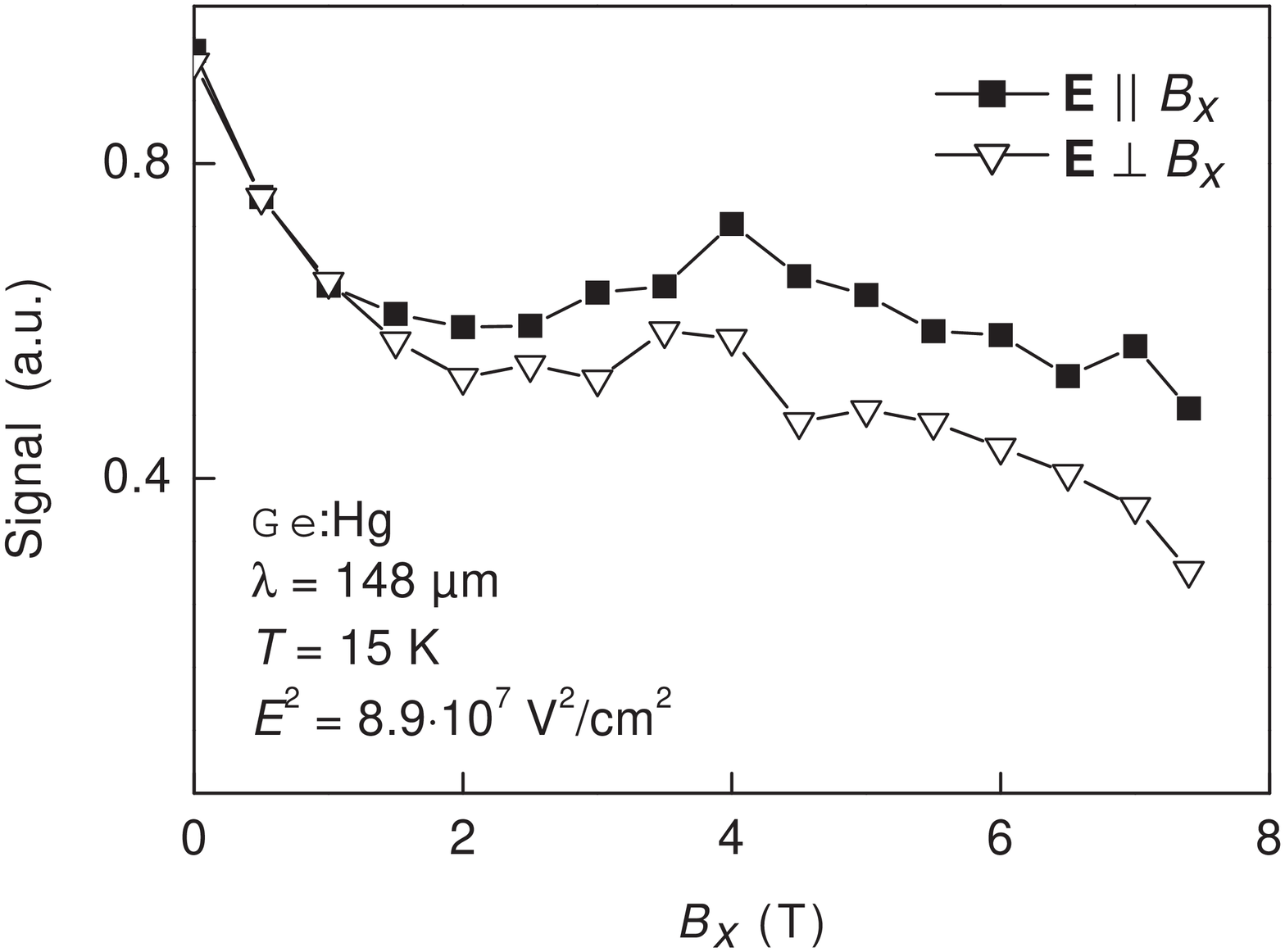}
\caption{Photoconductive signal as a function of magnetic field
strength for two polarizations: $\textbf{\textit{E}} \perp B_x$
and $\textbf{\textit{E}} \parallel B_x$ in the high-frequency
limit ($\omega\tau > 1$). } \label{F1}
\end{figure}

\begin{figure}
\centering
\includegraphics[scale=0.8]{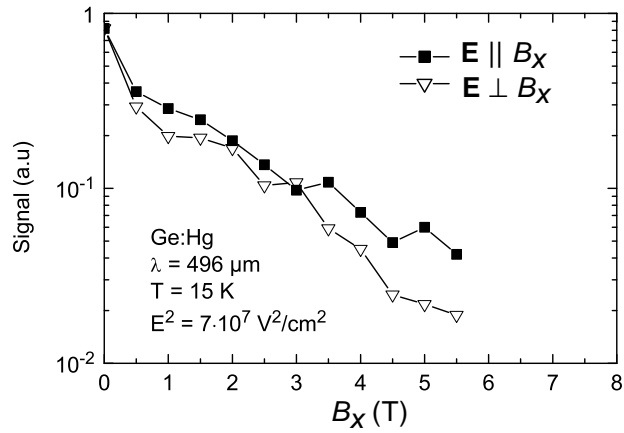}
\caption{Photoconductive signal as a function of magnetic field
strength for two polarizations: $\textbf{\textit{E}} \perp B_x$
and $\textbf{\textit{E}} \parallel B_x$ in the quasi-static limit
($\omega\tau<1$)
   }
\label{F3}
\end{figure}
%

In Fig.~2 the dependence of the photoconductive signal being
proportional to the ioni\-za\-tion probability on the square of
the electric field strength of the radiation is plotted for $B=0$
and $B=7.4$~T. Results are presented for a temperature of 15~K and
$\lambda=$ 148~$\mu$m. These measurements show that at zero
magnetic field the ionization probability is independent on the
electric field orientation. At high magnetic fields, however, the
signal for $\textit{\textbf{E}}\perp B_x$ drops significantly
below that of $\textbf{\textit{E}}\parallel B_x$. This suppression
of the tunneling probability can also be seen in the magnetic
field dependence of the photoconductive signal shown in Fig.~3 and
4. The effect of tunneling suppression occurs only at low
temperatures where, on the one side, practically all carriers are
frozen out on the impurities and on the other side the tunneling
time assumes high values~\cite{review2003tun}. At high
temperatures the effect of free carrier absorption interferes with
the change of conductivity by tunneling ionization. Therefore no
suppression of the signal could be  observed. Tunneling
suppression has been observed in both regimes, the quasi-static
limit ($\omega\tau<1$) at 496\,$\mu$m (Fig.~3) and in the
high-frequency limit  ($\omega\tau > 1$) obtained by the
excitation with radiation of  148\,$\mu$m (Fig.~4).

In summary,  our present observation shows that for
$\textbf{\textit{B}} \perp \textbf{\textit{E}}$ the magnetic field
deflects the carriers, which increases the length of the tunneling
trajectory. Thus, a magnetic field reduces the ionization
probability if the cyclotron frequency becomes larger than the
reciprocal tunneling time. Experimental findings are in good
agreement with the de\-ve\-loped theory of phonon-assisted
tunneling ionization of deep impurities in the presence of a
magnetic field.


\begin{thebibliography}{99}
%
\setlength{\itemsep}{-3pt}
\raggedright
%
\bibitem{review2003tun} S.D.\,Ganichev,
I.N.\,Yassievich, and W.\,Prettl, J. Phys.: Condens. Matter {\bf
14}, R1263-R1295 (2002)
\bibitem{Landauer1994p217}R. Landauer and
Th. Martin, Rev. Mod. Phys. {\bf 66}, 217 (1994).
\bibitem{Kotova68p616} L.P. Kotova, A.M. Perelomov, and V.S. Popov, Sov. JETP {\bf 27}, 616
(1968).
\bibitem{Eaves}L.~Eaves, K.W.H.~Stevens, and F.W.~Sheard,
in: {\em The physics and fabrication of microstructures and
microdevices}, editors M.J.~Kelly and C.~Weisbuch, p. 343-349,
Springer-Verlag, Berlin 1986.
\bibitem{Perel98p804}
V.I.Perel, I.N.Yassievich, JETP Lett. {\bf 68}, 804 (1998).
\bibitem{physicaB99} A.S. Moskalenko,  S.D.~Ganichev,  V.I. Perel, and
I.N. Yassievich, Physica B {\bf 273-274}, 1007 (1999).
\bibitem{Moskalenko00p217} A.S.~Moskalenko, V.I.~Perel, and I.N.~Yassievich, JETP {\bf 90}, 217
(2000).
\bibitem{PRL95} S.D.~Ganichev,  J.~Diener, I.N.~Yassievich, W.~Prettl,
B.K.~Meyer, and K.W.~Benz, Phys. Rev. Lett.~{\bf75}, 1590 (1995).
%
\end{thebibliography}
\end{document}